\documentclass[pra, 10pt, letterpaper,oneside,notitlepage, twocolumn]{revtex4-1}
\usepackage[latin1]{inputenc}
\usepackage{mathtools}
\usepackage{amsfonts}
\usepackage{array}
\usepackage[T1]{fontenc}
\usepackage{amssymb}
\usepackage[left=.75in, right=.75in, top=1in, bottom=1in]{geometry}
\usepackage{amssymb}
\usepackage{graphics}
\usepackage{graphicx}
\usepackage{array}
\usepackage{color}
\usepackage[hypertex, colorlinks, colorlinks=false, linkcolor=blue, citecolor=green, filecolor=black, urlcolor=cyan]{hyperref}

\begin{document}

\title{General solution to nonlinear optical quantum graphs using Dalgarno-Lewis summation techniques}
\author{Rick Lytel}
\author{Sean M. Mossman}
\author{Mark G. Kuzyk}
\affiliation{Department of Physics and Astronomy, Washington State University, Pullman, Washington  99164-2814}

\begin{abstract}
We develop an algorithm to apply the Dalgarno-Lewis (DL) perturbation theory to quantum graphs with multiple, connected edges. We use it to calculate the nonlinear optical hyperpolarizability tensors for graphs and show that it replicates the sum over states computations, but executes ten to fifty times faster.  DL requires only knowledge of the ground state of the graph, eliminating the requirement to determine all possible degeneracies of a complex network.  The algorithm is general and may be applied to any quantum graph.
\end{abstract}



\maketitle

\section{Introduction}\label{sec:start}
Quantum graphs were recently introduced as general models to explore the fundamental limits of nonlinear optics and to discover the topological and geometrical properties of structures that maximize the nonlinear optical response\cite{shafe12.01,lytel13.01,lytel13.04,lytel14.01,lytel15.01}.  Like other Hamiltonian models, quantum graphs with specific shapes and topologies have nonlinear optical response near the fundamental limits\cite{lytel16.01}.

A quantum graph is a metric graph on which electron dynamics is confined to the edges of the graph.  Quantum graphs were first studied as tractable molecular models\cite{pauli36.01,kuhn48.01,ruede53.01,scher53.01,platt53.01} and have been invoked as models of mesoscopic systems\cite{kowal90.01}, optical waveguides \cite{flesi87.01}, quantum wires\cite{ivche98.01,sanch98.01}, excitations in fractals \cite{avish92.01}, and fullerines, graphene, and carbon nanotubes\cite{amovi04.01,leys04.01,kuchm07.01}. Quantum graphs are also exactly solvable models of quantum chaos\cite{kotto97.01,kotto99.02,kotto00.01,blume01.04}.  In nonlinear optics, quantum graphs are models of a branched nano-wire structure, or a quasi-linear molecule, such as a donor-acceptor, with side groups.

The solution of quantum graphs for nonlinear optics has used the sum over states perturbation theory developed decades ago\cite{orr71.01}.  The process requires determination of the eigenfunctions and energies of the graph for large numbers of states, and careful determination of the degenerate states comprising the spectrum.  A Monte Carlo calculation of the response for a simple graph can take hours of computer time, frequent checks that the solutions for each graph are correct through the use of sum rules, and careful delineation of all degeneracies.  The value of such simulations to the molecular designer has been investigated, and new design rules have been posited\cite{lytel15.02}.  But the method does not scale well to graphs with many edges and large numbers of degenerate states that are often impossible to determine from a direct diagonalization of the Hamiltonian.

Recently, the Dalgarno-Lewis (DL) formulation of perturbation theory\cite{dalga55.01} has been adapted to off-resonance and dispersive nonlinear optics\cite{mossm15.01}.  The DL approach requires knowledge of only the ground state of the structure.  (For intrinsic nonlinearities\cite{kuzyk00.01}, the ground and first excited state energies are also required.)  As such, it offers a huge computational advantage for large-scale calculations, as only a single eigenfunction must be specified.  Degeneracies are implicitly incorporated into the DL formulation.  Application of the DL formalism to quantum graphs opens up the possibility of a general computational approach for exploring the theory of fundamental limits, as well as for simulating molecular and nanostructures that might achieve larger intrinsic nonlinear optical responses.

Section \ref{sec:QG} reviews the solution of quantum graphs and the use of the sum over states to compute the nonlinearities.  Section \ref{sec:DL} introduce the DL formalism and formally adapts it directly to quantum graphs.  Section \ref{sec:DLQG} develops the general algorithm for solving quantum graphs using the results of Section \ref{sec:DL} and illustrates it with wire, star, loop, and a composite, seven-edged graph, and displays the accuracy and computational speedup of the method over the sum over states.  Section \ref{sec:end} summarizes the benefits of the method and its extension to dispersive nonlinear optics.

\section{Computation with quantum graphs}\label{sec:QG}
Optimization studies using quantum graphs are based upon Monte Carlo computations with tens to hundreds of thousands of possible geometries for a particular topology, such as a star or loop motif\cite{lytel13.02}, or a composite of stars, loops, and wires.  For each structure, the Hamiltonian is diagonalized to find the energy spectrum $E_{n}$ and the eigenstates $|n\rangle$, where $n$ is the mode number.  The transition moments $r^{i}_{nm}=\langle n|r^{i}|m\rangle$ are computed for the graph using a union operation\cite{lytel13.01}, and the first and second hyperpolarizability tensors $\beta_{ijk}$ and $\gamma_{ijkl}$ are calculated using perturbation theory based upon the sum over states\cite{orr71.01}:
\begin{equation}\label{beta_ijk}
\beta_{ijk}=\left(\frac{e^3}{2}\right)P_{ijk}{\sum_{nm}}'\frac{r_{0n}^{i}\bar{r}_{nm}^{j}r_{m0}^{k}}{E_{n0}E_{m0}}
\end{equation}
and
\begin{eqnarray}\label{gamma_ijkl}
\gamma_{ijkl}&=&\left(\frac{e^4}{6}\right)P_{ijkl}\Biggl({\sum_{nmp}}'\frac{r_{0n}^{i}\bar{r}_{nm}^{j}\bar{r}_{mp}^{k}r_{p0}^{l}}{E_{n0}E_{m0}E_{p0}}\nonumber \\
&-&{\sum_{nm}}\frac{'r_{0n}^{i}r_{n0}^{j}r_{0m}^{k}r_{m0}^{l}}{E_{n0}E_{n0}E_{m0}}\Biggr)
\end{eqnarray}
where $\bar{r}^{i}_{nm}=r^{i}_{nm}-\delta_{nm}r^{i}_{00}$, $E_{n0}=E_{n}-E_{0}$, and the permutation operator P permutes all indices into all possible permutations without regard to order.  The indices are $i=x,y$ for a planar graph embedded in two-dimensional space.  Both hyperpolarizabilities have been normalized to their maximum values\cite{kuzyk00.01} and are bounded above by unity.  The lower bound on $|\beta_{ijk}|$ is zero, while that for $\gamma_{ijkl}$ is $-0.25$.  The geometry of the structure largely determines the total contributions of the transition moments from each edge, each of which is determined by the mode overlap of the eigenfunctions with the appropriate position operator and the relative orientation of the edges.  Tailoring this overlap and the change in the dipole moment for electron excitation from the ground state to an excited state can lead to a giant enhancement of the nonlinear optical response due to phase disruption of the eigenfunctions $\psi_{n}(s)=\langle s|n\rangle$, where $s$ is measured along the graph\cite{lytel15.02}.

\begin{figure}\centering
\includegraphics[width=3.4in]{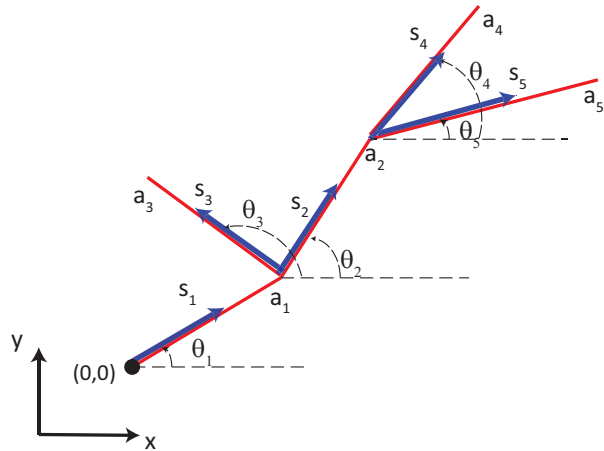}\\
\caption{A five-edge quantum graph, with two 3-star motifs, defining the coordinates and geometry used in the text.}\label{fig:graph5edge}
\end{figure}

Figure \ref{fig:graph5edge} illustrates the notation for computation with quantum graphs.  Each edge has a coordinate $s_{p}$ measured along the edge from zero to $a_{p}$ and an angle $\theta_{p}$ with respect to an external axis, which we choose to be the $x-axis$ for convenience.  The edge functions $\phi^{(p)}_{n}(s_{p})$ for edge $p$ and mode $n$ on that edge satisfy $H\phi^{(p)}_{n}=E_{n}\phi^{(p)}_{n}$, where the energies $E_{n}$ are the same on all $p$ edges.  The full eigenfunctions $\psi_{n}$ are constructed using a mathematical union operation
\begin{equation}\label{edgeUnion}
\psi_{n}(s)=\bigcup_{p=1}^{E}\phi^{(p)}_{n}(s_{p})
\end{equation}
operating on all $E$ edges\cite{lytel13.01}.  The Hilbert space for the solutions to the entire graph is a direct sum of those for each individual edge.  The interpretation of the coordinates in Eqn \ref{edgeUnion} is that when the electron is on edge $p$, the eigenfunction $\psi_{n}(s)$ is equal to $\phi_{n}^{(p)}(s_{})$ for any mode $n$.

The construction of the union of edge functions to make an eigenfunction of the graph requires matching of boundary conditions at the vertices within the graph.  For quantum graph unions, the boundary conditions are that the edge functions at each vertex are equal, and that the net probability flux into or out of a vertex is zero, or equivalently
\begin{equation}\label{flux}
\sum_{p}^{E_{v}}\frac{d\phi^{(p)}_{n}(s_{p})}{ds_{p}}\Big|_{v}=0
\end{equation}
where the sum is over the $E_{v}$ edges going into or out of the vertex $v$.  There are always enough boundary conditions to compute all of the constants of integration from the solutions of the edge functions.

The transition moments are computed by projecting the coordinate vector $s_{i}$ for a given edge $i$ onto the $x-axis$ to compute $\langle n|x^{i}|m\rangle =\langle n|s_{i}|m\rangle \cos{\theta_{i}}$ and onto the $y-axis$ to compute $\langle n|y^{i}|m\rangle =\langle n|s_{i}|m\rangle \sin{\theta_{i}}$ for each edge. Projected into coordinate space, the transition moments take the form of sums of definite integrals.  For example, for $x_{nm}$, we get
\begin{equation}\label{moments}
x_{nm}=\sum_{p=1}^{E}\cos{\theta_{p}}\int_{0}^{a_{p}} ds_{p}s_{p}\phi^{(p)}_{n}(s_{p})\phi^{(p)}_{m}(s_{p})
\end{equation}
where the sum is over the $E$ edges, with a similar expression with $cos\rightarrow sin$ for $y_{nm}$.  The solution of a quantum graph yields the spectrum and transition moments required by Eqns \ref{beta_ijk} and \ref{gamma_ijkl} for the hyperpolarizability tensors.  Once a graph is specified by its metric variables $a_{p},\theta_{p}$ and the potentials acting on each edge, the dependence of the magnitudes of the hyperpolarizabilities on the geometry and topology of the graph is computable and may be studied using Cartesian or spherical tensor analysis\cite{jerph78.01,lytel13.01}.

Explorations of the topological structures maximizing the hyperpolarizabilities require computation of $\beta_{ijk}$ and $\gamma_{ijkl}$ for each of tens of thousands of graphs.  Computation for each graph requires solving the graph for its energies and eigenfunctions, typically for at least $10-50$ modes, depending on the nature of the potential energy.  A single Monte Carlo run for a simple three-edged graph might require several hours of execution time on an eight-threaded, quad-core Intel $i7$ CPU clocked at $3.7 GHz$.  The computation requires careful handling of the solutions of the graph to ensure that any and all degenerate states are discovered and enumerated.  For a typical computation, it is necessary to invoke the Thomas-Reiche-Kuhn (TRK) sum rules\cite{thom25.01,reich25.01,kuhn25.01} to ensure that all such states have been correctly captured and the spectra correctly determined\cite{lytel13.01}.  Computation of a quantum graph for nonlinear optics is an elaborate procedure that is time-consuming and often frustrating when graph geometries produce numerous degeneracies.  The Dalgarno-Lewis perturbation theory offers a simplification of the Monte Carlo computation and an execution time speedup of over a factor of $50$.

\section{Dalgarno-Lewis perturbation theory}\label{sec:DL}
The Dalgarno-Lewis perturbation method has been reviewed in the literature\cite{mavro91.01,dalga55.01,harri77.01,maize11.01} and in classic quantum mechanics textbooks\cite{schif68.01}.  A paper in the present volume describes its application to dispersive nonlinear optics and solves a number of one-dimensional examples.  Quantum graphs are quasi-one dimensional objects, with particle motion along the edges of the graph and not transverse to it.  But they live in two dimensions and have $x$ and $y$ projections that contribute to the hyperpolarizability tensors.  Moreover, their Hilbert space is a direct sum of Hilbert spaces for each of the edges, which adds complexity to the DL formalism and for which no algorithms have yet been developed.  This section summaries the DL formalism for one dimension and then extends it to quantum graphs for the first time.

\subsection{DL theory in one dimension}\label{sec:DLone}
The Dalgarno-Lewis method for perturbation theory in one dimension begins by assuming the existence of an operator $F$ whose commutator with the Hamiltonian $H=p^2/2m+V$ takes the form
\begin{equation}\label{defineF}
\left[F,H\right]=\bar{x}
\end{equation}
with $\bar{x}=x-x_{00}$. The right-hand side is essentially the perturbing potential on a molecule in an electric field, $V=-ex$, with the ground state expectation value subtracted and the electric charge e dropped, as it will be incorporated elsewhere later.

Acting on the ground state with both sides of Eqn \ref{commFH} yields
\begin{equation}\label{commFH}
\left[F,H\right]|0\rangle = \bar{x}|0\rangle
\end{equation}
If we operate on the left of both sides of Eqn \ref{commFH} with $\langle n|$ and use $H|n\rangle = E_{n}|n\rangle$, we get
\begin{equation}\label{Fn0}
F_{n0}=-\frac{\bar{x}_{n0}}{E_{n0}}
\end{equation}
which defines matrix elements of the F operator in a way that allows a substitution into the sum over states expressions for $\beta_{xxx}$:
\begin{eqnarray}\label{betaxxx}
\beta_{xxx}&=&3e^3{\sum_{n,m}}'\frac{x_{0n}\bar{x}_{nm}x_{m0}}{E_{n0}E_{m0}} \\
&=&-3e^3{\sum_{n,m}}'F_{0n}\bar{x}_{nm}F_{m0}\nonumber \\
&=& -3e^3\langle 0|F\bar{x}F|0\rangle\nonumber
\end{eqnarray}
where the last line in Eqn \ref{betaxxx} follows by using completeness of the eigenstates.  If the DL operator exists, then the first hyperpolarizability is computable through the ground state expectation value of the composite operator $F\bar{x}F$. Evidently, the effect of the perturbation on the excited states and their contributions to $\beta_{xxx}$ are embodied in the $F$ operator.

To obtain a spatial representation for $F$ that we may use in computations, we project Eqn \ref{commFH} into coordinate space as follows.  The Hamiltonian, in coordinate space, is given by
\begin{equation}\label{Hamiltonian}
H=-\frac{\hbar^2}{2m}\frac{d^2}{dx^2} + V(x),
\end{equation}
We want to project Eqn \ref{defineF} into position space, where we know $\langle x|0\rangle=\psi_{0}(x)$ is the ground state wavefunction.  We will need the projection of F into x-space, which is defined through
\begin{eqnarray}\label{F0}
\langle x|F|0\rangle&=&\int dx'\langle x|F|x'\rangle\langle x'|0\rangle \\
&=&F(x)\psi_{0}(x)
\end{eqnarray}
which defines the function $F(x)$.  We are asserting that $F(x)$ exists for our operator.  If we can derive an expression for it, then we will have found a representation for the $F$ operator and may use Eqn \ref{betaxxx} to compute the first hyperpolarizability as an expectation value of $F\bar{x}F$ in the \emph{unperturbed} ground state.  Note that we do not even need to know the spectrum.  The energy scale won't enter into the problem until we divide by $\beta_{max}$ from the theory of fundamental limits.  We'll return to this later.

Projecting the left-hand side (LHS) of Eqn \ref{commFH} into x-coordinate space gives
\begin{eqnarray}\label{LHSF}
LHS&=& F(x) \left(- \frac{\hbar^2}{2m} \frac{d^2}{dx^2} + V(x)\right) \psi_{0}(x) \\
&-&\left( - \frac{\hbar^2}{2m} \frac{d^2}{dx^2} + V(x) \right) F(x) \psi_{0}(x)\nonumber \\
&=& \frac{\hbar^2}{2m} \left(\frac{d^2F(x)}{dx^2}\psi_{0}(x) + 2 \frac{dF(x)}{dx} \frac{d\psi_{0}(x)}{dx}\right)\nonumber \\
&=&\frac{\hbar^2}{2m}\frac{1}{\psi_{0}(x)} \frac{d}{dx}\left(\left(\psi_{0}(x)\right)^2\frac{dF(x)}{dx}\right)\nonumber.
\end{eqnarray}
Projecting the right-hand side of Eqn \ref{commFH} into coordinate space yields
\begin{equation}\label{RHSF}
RHS=\bar{x}\psi_{0}(x)
\end{equation}
Equating Eqns \ref{LHSF}, and \ref{RHSF}, we get the differential equation
\begin{equation}\label{diffEQforF}
\frac{d}{dx}\left(\left(\psi_{0}(x)\right)^2\frac{dF(x)}{dx}\right) = \frac{2m}{\hbar^2}\bar{x} \left(\psi_{0}(x)\right)^2.
\end{equation}
The general solution to Eqn \ref{diffEQforF} is obtained with two quadratures:
\begin{eqnarray}\label{Fofx}
F(x)&=&F(0)+\frac{2m}{\hbar^2}\int_{0}^{x}\frac{dx'}{\psi_{0}(x')^{2}}\\
&\times&\left(\int_{0}^{x'}dx''{\bar{x}}''\psi_{0}(x'')^{2}+\frac{\hbar^2}{2m}F'(0)\psi_{0}(0)^{2}\right)\nonumber
\end{eqnarray}

Eqn \ref{Fofx} is the main result.  Note that the lower limit of the inner integral could have been chosen to be some value $\delta$, where $F'(\delta)\psi_{0}(\delta)^2=0$, but we choose $\delta=0$ for quantum graphs.  Once the ground state wavefunction is known, the function $F(x)$ may be found by analytical or numerical integration, once two boundary conditions are specified.  Actually, for bounded problems, such as a quantum well where the wavefunction vanishes at each end, the second term in Eqn \ref{Fofx} vanishes without specifying a boundary condition $F'(0)$. We may always choose $F(0)$ any way we please, because the operator $F$ is determined by a commutator and is therefore only known up to a constant anyway.

To calculate $\gamma$ from the sum over states, we need an additional operator $G$.  In analogy with Eqn \ref{commFH}, we intuit that this is
\begin{equation}\label{commGH}
\left[G,H\right]|0\rangle = \Big[\bar{x}F-\langle 0|\bar{x}F|0\rangle\Big] |0\rangle
\end{equation}
which leads to a differential equation for $G(x)$, the projection of $G$ into coordinate space.  The solution to this equation may be written down by inspection of Eqn \ref{Fofx} and is
\begin{eqnarray}\label{Gofx}
G(x)&=& G(0) \\
&+&\frac{2m}{\hbar^2}\int_0^x \frac{dx'}{\psi_{0}(x')^{2}}\Bigg( \int_{0}^{x'}dx''\big[{\bar{x}}''F(x'')\nonumber \\
&-&<0|{\bar{x}}''F|0>\big] \psi_{0}(x'')^{2} +\frac{\hbar^2}{2m}G'(0)\psi_{0}(0)^{2}\Bigg).\nonumber
\end{eqnarray}
Eqn \ref{commGH} implies that
\begin{equation}\label{Gn0}
G_{n0}=-\frac{\langle\bar{x}F\rangle_{n0}}{E_{n0}}
\end{equation}
which may be used to write Eqn \ref{gamma_ijkl} in a compact sum of ground state expectation values of operator products of the form $F^px^qG^r$ with integers $p,q,r$. We present an explicit form in the next section that is applicable to quantum graphs.

Note that we have defined the action of the anti-Hermitian operator $F$ on the ground state $F|0\rangle$ to be equal in x-space to $F(x)\psi_{0}(x)$.  This implies that the adjoint has a minus sign:  $\langle 0|F$ becomes $-F^{*}(x)\psi_{0}^{*}(x)$.  This is important because operator products like $FF$ will produce minus signs when they are sandwiched between ground states.  The same is true for $G$ if it acts to the left.

\subsection{Extension to quantum graphs}\label{sec:exten}
We can easily extended the one dimensional Dalgarno-Lewis formalism to quasi-one dimensional graphs using the fact that the Hamiltonian rotates as a quadratic form, viz., $H(x,y)\rightarrow H(s)+H(\tau)$ and ignoring the transverse part\cite{shafe11.02}.  This implies that all of our one-dimensional work carries through for the $F$ and $G$ functions, except now the coordinates on the right hand side of Eqn \ref{commFH} and \ref{commGH} are either $x$ or $y$, depending on the transition moment one wishes to calculate.  With this in mind, we arrive at
\begin{eqnarray}\label{Fi}
F_{i}(s)&=& F_{i}(0)+\frac{2m}{\hbar^2}\int_0^s \frac{ds'}{\psi_{0}(s')^{2}} \\
&\times&\Bigg( \int_{0}^{s'}\bar{r}^{i}(s'')\psi_{0}(s'')^{2} ds''+\frac{\hbar^2}{2m}{F_{i}}'(0)\psi_{0}(0)^{2}\Bigg)\nonumber
\end{eqnarray}
where $r^{1}=x$ and $r^{2}=y$. The coordinate variable $s$ is measured along the graph edges.  The wavefunction and DL functions in Eqn \ref{Fi} are unions of functions defined on the graph edges.  The interpretation of the integral is that when $s$ is on edge $p$, the values of the edge functions on edge $p$ are to be used, and the coordinate variable spans the range of that edge function.  We quantify this description in Section \ref{sec:DLQG}.

For $\gamma$, we use the new set of functions, $G_{ij}$ defined by the commutator $[G_{ij},H]=\bar{r}^{i}F_{j}-<0|\bar{r}^{i}F_{j}|0>$, which translates in s-space to an equation for $G_{ij}(s)$ which is identical to that for $F_{i}$ in Eqn \ref{Fi}, but with an inner integrand given by the product $\bar{r}^{i}F_{j}-<0|\bar{r}^{i}F_{j}|0>$ expressed in s-space.  The explicit form is
\begin{eqnarray}\label{Gij}
G_{ij}(s)&=& G_{ij}(0)=\frac{2m}{\hbar^2}\int_0^s \frac{ds'}{\psi_{0}(s')^{2}} \\
&\times&\Bigg( \int_{0}^{s'}\big[\bar{r}^{i}(s'')F_{j}(s'')-<0|\bar{r}^{i}F_{j}|0>\big] \psi_{0}(s'')^{2} ds''\nonumber \\
&+&\frac{\hbar^2}{2m}{G_{ij}(0)}'\psi_{0}(0)^{2}\Bigg).\nonumber
\end{eqnarray}
Note that the $s$ variable appearing in each of Eqns \ref{Fi} and \ref{Gij} roams over the entire graph and that $\psi_{0}(s)$ is the ground state wavefunction for the graph, which is the union of the edge functions.

It is now straightforward to derive simple expressions for the hyperpolarizabilities.  First, we note that it is always possible to add a constant to F to make $\langle 0|F|0\rangle=0$ so we will assume that has always been done.  Eqn \ref{Fn0} then allows us to write the expression for $\beta_{ijk}$ in Eqn \ref{beta_ijk} without energy denominators as
\begin{equation}\label{beta_ijk_simple}
\beta_{ijk}=-P_{ijk}\left(\frac{e^3}{2}\right)\left(\langle 0|F_{i}r^{j}F_{k}|0\rangle-r_{00}^{j}\langle 0|F_{i}F_{k}|0\rangle\right).
\end{equation}
We note that we can also add a constant to G to make $\langle 0|G|0\rangle=0$ without changing the commutator that defined G.  We will always assume we have done this.  Given this, we may use Eqn \ref{Gn0} to write down the general expressions for $\gamma_{ijkl}=\gamma^{(1)}_{ijkl}-\gamma^{(2)}_{ijkl}$ as
\begin{eqnarray}\label{gamma_ijkl_simple}
\gamma^{(1)}_{ijkl}&=&P_{ijkl}\left(\frac{e^4}{6}\right)\langle 0|F_{i}\bar{r}^{j}G_{kl}|0\rangle \\
\gamma^{(2)}_{ijkl}&=&P_{ijkl}\left(\frac{e^4}{6}\right)\langle 0|F_{i}F_{j}|0\rangle\langle 0|r^{k}F_{l}|0\rangle\nonumber
\end{eqnarray}

This completes the \emph{formal} specification of the F and G functions, and their use in calculating the hyperpolarizabilities.  To apply this formalism to quantum graphs, we must express the DL functions in the language of the direct sum Hilbert space they represent.

\section{General solution for quantum graphs.}\label{sec:DLQG}
For graphs, the meaning of eigenfunctions $\psi_{n}(s)$ and ground state expectation values require sharp definition of the union operation that stitches together the Hilbert spaces on each edge to yield a direct sum Hilbert space with energies equal to those of the spaces for each edge.  Continuity and flux conservation, Eqn \ref{flux}, showed how this is accomplished for the edge functions.  With this algorithm, Eqn \ref{moments} showed how expectation values of operators acting across a graph are calculated, edge by edge, for transition moments.  For $F_{i}$ and $G_{ij}$, we need the union of an $F$ or $G$ function tied to each edge:
\begin{equation}\label{Fip}
F_{i}(s)=\bigcup_{p=1}^{E}F_{i}^{(p)}(s_{p})
\end{equation}
with a similar union operation defined for $G_{ij}$:
\begin{equation}\label{Fip}
G_{ij}(s)=\bigcup_{p=1}^{E}G_{ij}^{(p)}(s_{p}).
\end{equation}
Here and going forward, the index $i$ refers to the Cartesian component $x$ or $y$ of an object, while the index $p$ refers to a mode number.

The matching of the edge $F_{i}^{(p)}$ and $G_{ij}^{(p)}$ functions at vertices is accomplished by demanding continuity of each at the vertex and a net flow into or out of the vertex equal to zero.  A ground state expectation value of the form of Eqn \ref{moments} (with m=n=0) but for $xF_{x}$, say, would then be calculated as
\begin{equation}\label{xF00}
\langle 0|xF_{x}|0\rangle=\sum_{p=1}^{E}\cos{\theta_{p}}\int_{0}^{a_{p}} ds_{p}s_{p}F_{x}^{(p)}(s_{p})\phi_{0}^{(p)}(s_{p})\phi_{0}^{(p)}(s_{p})
\end{equation}
with a similar expression for expectation values of other products of operators.  The boundary conditions arising from the union operation completely specify all integration constants appearing in the edge DL functions and provide a general solution to the problem of calculating the first and second hyperpolarizabilities from only expectation values of products of DL operators.  For clarity, going forward, \emph{we drop the mode subscript $0$ on the edge functions in the ground state by writing $\phi^{p}_{0}(s_{p})\equiv\phi^{p}(s_{p})$}.

The explicit forms of the $F_{i}^{(p)}$ and $G_{ij}^{(p)}$ DL edge functions are
\begin{eqnarray}\label{Fip}
F_{i}^{(p)}(s_{p})&=& F_{i}^{(p)}(0)+\frac{2m}{\hbar^2}\int_0^{s_{p}} \frac{ds_{p}'}{\phi^{(p)}(s_{p}')^2} \\
&\times&\Bigg( \int_{0}^{s_{p}'}\bar{r}^{i}(s_{p}'')\phi^{(p)}(s_{p}'')^2 ds_{p}''+\frac{\hbar^2}{2m}{F'_{i}}^{(p)}(0){\phi^{(p)}}(0)^2\Bigg)\nonumber
\end{eqnarray}
and
\begin{eqnarray}\label{Gijp}
G_{ij}^{(p)}(s_{p})&=& G_{ij}^{(p)}(0)=\frac{2m}{\hbar^2}\int_0^{s_{p}} \frac{ds_{p}'}{\phi^{(p)}(s_{p}')^2} \\
&\times&\Bigg( \int_{0}^{s_{p}'}[\bar{r}^{i}(s_{p}'')F_{j}^{(p)}(s_{p}'')-\langle 0|\bar{r}^{i}F_{j}|0\rangle] \phi^{(p)}(s_{p}'')^2 ds''\nonumber \\
&+&\frac{\hbar^2}{2m}{G'_{ij}}^{(p)}(0){\phi^{(p)}(0)}^2\Bigg).\nonumber
\end{eqnarray}
where the angular brackets signify the expectation value in the ground state DL edge function.  The integration constants are selected to satisfy the boundary conditions on the DL functions so their union is a function that is the projection of the DL operator $F$ or $G$ into coordinate space over the entire graph.  The boundary conditions are that the $F_{i}^{(p)}(s_{p})$ are equal at the vertex where the edges which they are index meet, and the sum of the derivatives with respect to their edge coordinates, measured into or out of a vertex, of all edges meeting at that vertex vanishes.  Similar conditions hold for $G_{ij}^{(p)}$ and its derivatives.

The first integral of the DL equation for $F_{i}^{(p)}(s_{p})$ for any edge p of a graph may be written as
\begin{equation}\label{Fp1}
{F'_{x}}^{(p)}(s_{p})\phi^{(p)}(s_{p})^2={F'_{x}}^{(p)}(0)\phi^{(p)}(0)^2+\frac{2m}{\hbar^2}\int_{0}^{s_{p}}ds'_{p}\bar{x}(s'_{p})\phi^{(p)}(s'_{p})^2
\end{equation}
where $\bar{x}(s'_{p})=x_{0p}+\cos{\theta_{p}}s'_{p}-x_{00}$ for the ith edge.  Here, $x_{0p}$ is the offset from the origin for edge p, and $F'=d/ds$.  In particular, the following relationship holds at the endpoints of an edge:
\begin{eqnarray}\label{Fp2}
&&{F'_{x}}^{(p)}(a_{p})\phi^{p}(a_{p})^2-{F'_{x}}^{(p)}(0)\phi^{(p)}(0)^2 \\
&=&\frac{2m}{\hbar^2}\int_{0}^{a_{p}}ds'_{p}\bar{x}(s'_{p})\phi^{(p)}(s'_{p})^2\nonumber \\
&\equiv& \frac{2m}{\hbar^2}\langle\bar{x}_{p}\rangle\nonumber
\end{eqnarray}
Note that each DL function for edge $p$ has two integration constants. We show how the boundary conditions determine these constants.  Write Eqn \ref{Fip} as
\begin{eqnarray}\label{Fi2}
{F_{x}}^{(p)}(s_{p})&=&{F_{x}}^{(p)}(0)+\frac{2me}{\hbar^2}\int_{0}^{s_{p}}\frac{ds'_{p}}{\phi^{(p)}(s'_{p})^2}\\
&\times&\left(\int_{0}^{s'_{p}}ds_{p}''\bar{x}(s''_{p})\phi^{(p)}(s''_{p})+C_{p}\right)\nonumber
\end{eqnarray}
where the first integration constants are
\begin{equation}\label{Cp}
C_{p}=\frac{\hbar^2}{2m}{F'_{x}}^{(p)}(0)\phi^{p}(0)^2
\end{equation}
and are proportional to the net flux entering edge $p$.  Now write Eqn \ref{Fp2} as
\begin{equation}\label{Cout}
{F'_{x}}^{(p)}(a_{p})\phi^{p}(a_{p})^2=\frac{2m}{\hbar^2}\big[C_{p}+\langle\bar{x}_{p}\rangle\big]
\end{equation}
which is proportional to the net flux exiting edge $p$ times the square of the $pth$ edge function evaluated at the exit vertex.  Eqns \ref{Cp} and \ref{Cout} reveal that two edges $p1$ and $p2$ have a common vertex, say at $a_{p1}=a_{p2}=0$, then the $C_{p1}$ and $C_{p2}$ for edges $p1$ and $p2$ are related by flux conservation.  Eqns \ref{Cp} and \ref{Cout} may thus be used to conserve DL function flux at each vertex. A final set of equations matches the values of the DL functions at each vertex:
\begin{equation}\label{Fbc}
{F_{x}}^{(E_1)}(v;1)={F_{x}}^{(E_2)}(v;2)=\ldots{F_{x}}^{(E_d)}(v;d)
\end{equation}
for every set of $d$ edges that come together at vertex $v$, and the notation $(v;n)$ means evaluate at the position $s_{n}$ for that edge at vertex $v$.

The continuity of $F_{x}^{(p)}$ and conservation of ${F'_{x}}^{(p)}$ resemble the boundary conditions on the edge functions in a quantum graph.  These embody the definition of a union of functions operating on the edges.

For the G functions, a similar approach applies, with Eqn \ref{Gijp} replaced by
\begin{eqnarray}\label{Gp1}
{G_{xx}}^{(p)}(s_{p})&=&{G_{xx}}^{(p)}(0)+\frac{2m}{\hbar^2}\int_{0}^{s_{p}}\frac{ds'_{p}}{\phi^{(p)}(s'_{p})^2}\\
&\times&\bigg(\int_{0}^{s'_{p}}ds_{p}''\left[\bar{x}(s''_{p}){F_{x}}^{(p)}(s''_{p})-\langle 0|\bar{x}F_{x}|0\rangle\right]\phi^{(p)}(s''_{p})^2\nonumber \\
&+& D_{p}\bigg)\nonumber
\end{eqnarray}
for the xx component, with suitable modifications made for the yy and off-diagonal components.  The constants $D_{p}$ are proportional to the $G_{xx}^{(p)}$ flux entering edge $p$.
\begin{equation}\label{Dp}
D_{p}=\frac{\hbar^2}{2m}{G'_{xx}}^{(p)}(0)\phi^{p}(0)^2
\end{equation}
where this latter derivative may be obtained from an expression similar to Eqn \ref{Fp2}, viz.,
\begin{eqnarray}\label{Gp2}
&& {G'_{xx}}^{(p)}(a_{p})\phi^{p}(a_{p})^2-{G'_{xx}}^{(p)}(0)\phi^{(p)}(0)^2 \\
&=&\frac{2m}{\hbar^2}\int_{0}^{a_{p}}ds'_{p}\big[\bar{x}_{p}(s'_{p})F_{x}^{(p)}(s'_{p})-\langle 0|\bar{x}F_{x}|0\rangle\big]\phi^{(p)}(s'_{p})^2\nonumber \\
&\equiv& \frac{2m}{\hbar^2}\langle\big[\bar{x}_{p}F_{x}^{(p)}-\langle 0|\bar{x}F_{x}|0\rangle\big]\rangle\nonumber
\end{eqnarray}
leading to
\begin{equation}\label{Dout}
{G'_{xx}}^{(p)}(a_{p})\phi^{(p)}(a_{p})^2=\frac{2m}{\hbar^2}\bigg[D_{p}+\langle\big[\bar{x}_{p}F_{x}^{(p)}-\langle 0|\bar{x}F_{x}|0\rangle\big]\rangle\bigg]
\end{equation}
where again the brackets on the last term are the ground state expectation value in edge $p$.

Eqns \ref{Dp} and \ref{Dout} may be used to conserve DL function flux at each vertex. A final set of equations matches the values of the DL functions at each vertex:
\begin{equation}\label{Gbc}
{G_{xx}}^{(E_1)}(v;1)={G_{xx}}^{(E_2)}(v;2)=\ldots{G_{xx}}^{(E_d)}(v;d)
\end{equation}
for every set of $d$ edges that come together at vertex $v$, where again the notation $(v;n)$ means evaluate at the position $s_{n}$ for that edge at vertex $v$.

The method derived in this section is applicable to any quantum graph.  It is easy to use and will be illustrated with examples of wires, stars, and loops, and then applied to solve a seven-edge graph with three stars.  This graph may have numerous degeneracies, depending on the relative edge lengths, that are challenging to discover\cite{lytel14.01,lytel15.01}.  With the DL method, the solutions may be written down by inspection and easily coded, requiring only the ground state of the graph.  This is an enormous simplification in solving the graph and leads to reduction in execution time for a large scale Monte-Carlo calculation from several hours to several minutes.

In the following examples, we consider $i=x$ for clarity.  A similar set of results applies for the y direction by replacing x with y in the inner integrals' transition moment on each edge. \emph{We suppress the x-index to simplify notation}.

\subsection{Wire graphs.}\label{sec:3graphs}
\begin{figure}\centering
\includegraphics[width=1.2in]{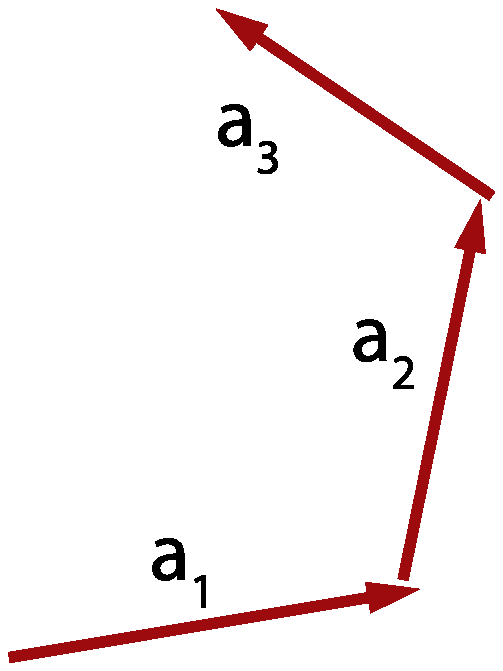}\includegraphics[width=1.1in]{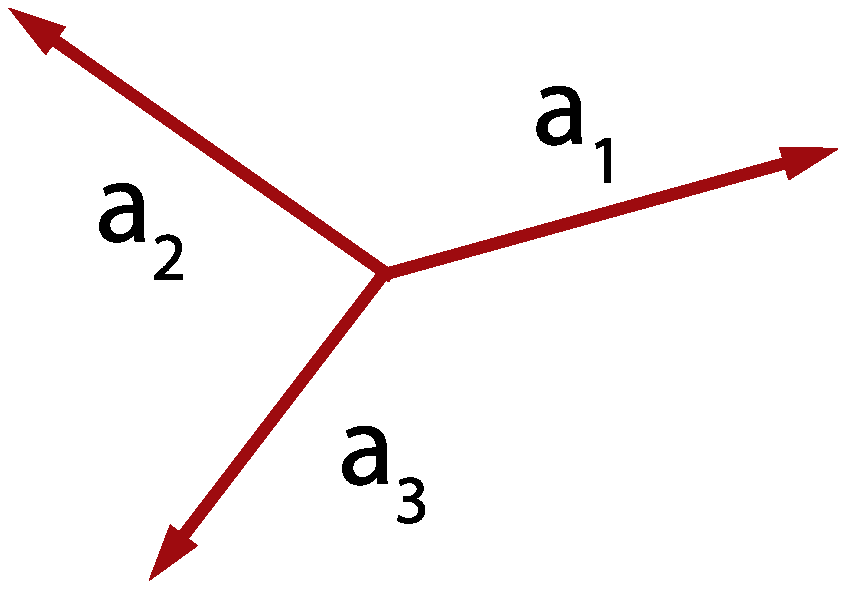}\includegraphics[width=0.9in]{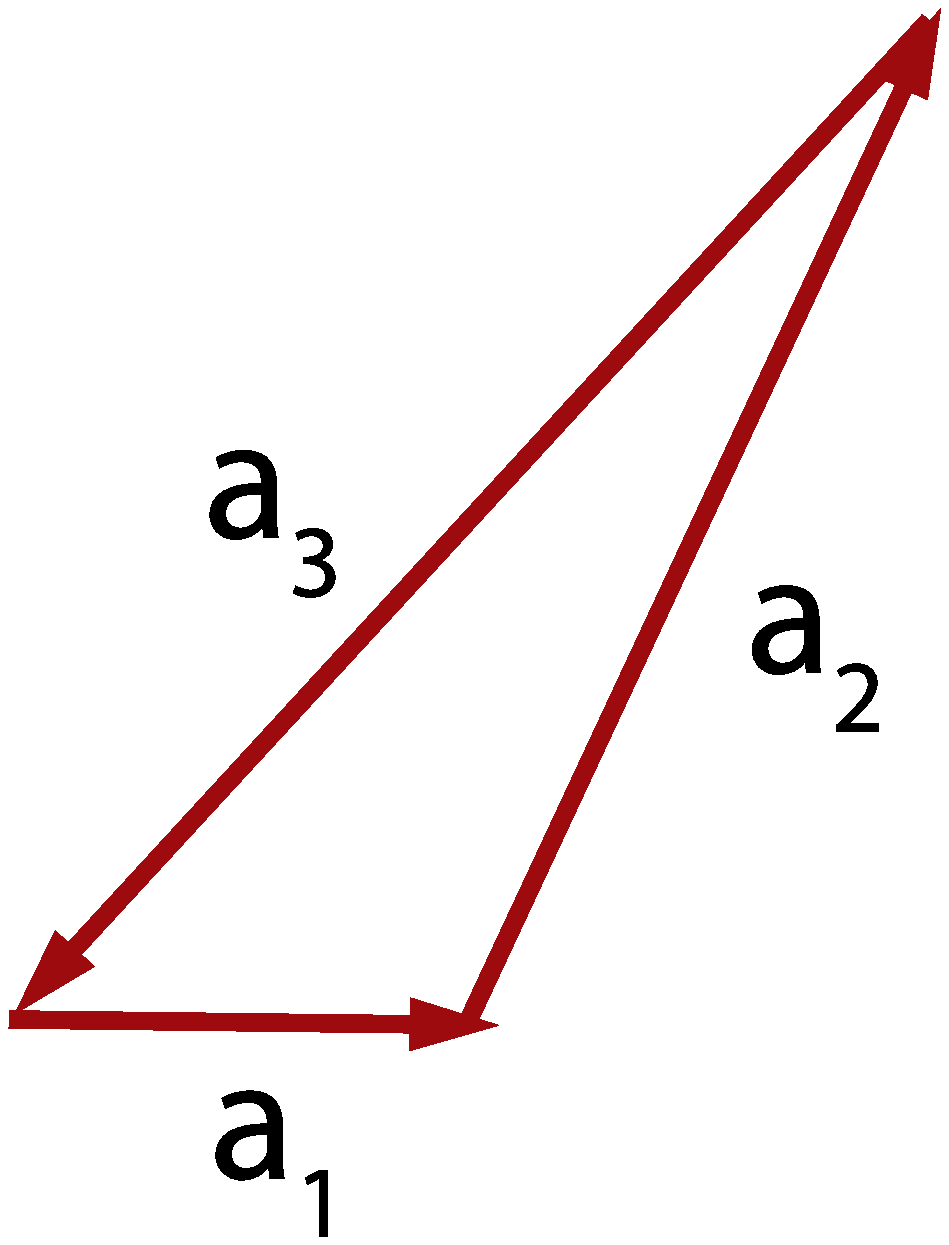}\\
\caption{Three wire (left), 3-star (center), and loop (right) graphs used to illustrate how to compute the F functions for the edges.}\label{fig:3graphs}
\end{figure}

Consider a three wire graph, as shown in the left panel of Figure \ref{fig:3graphs}.  We need to solve for the $F_{x}^{(p)}$ (and $F_{y}^{(p)}$) on each edge so that we may compute matrix elements of the union over edges of $F_{x}=\bigcup_{p}F_{x}^{(p)}$ (and $F_{y}=\bigcup_{p}F_{y}^{(p)}$) in the ground state.

Examining the graph and Eqn \ref{Cp}, it is easy to see that $C_{1}=0$ because the edge function $\phi^{(1)}(0)=0$.  To calculate $C_{2}$, we note that it is just the flux entering edge 2, which is also the flux exiting edge 1, and by Eqn \ref{Cout}, we get $C_{2}=C_{1}+\langle\bar{x}_{1}\rangle$, so $C_{2}=\langle\bar{x}_{1}\rangle$.  Continuing, the flux entering edge three is $C_{3}$ is the flux exiting edge 2, which by Eqn \ref{Cout} is $C_{2}+\langle\bar{x}_{2}\rangle$, so $C_{3}=\langle\bar{x}_{1}\rangle+\langle\bar{x}_{2}\rangle$. Summarizing,
\begin{eqnarray}\label{Ci3wire}
C_{1}&=&0 \\
C_{2}&=&\langle\bar{x}_{1}\rangle\nonumber \\
C_{3}&=&\langle\bar{x}_{1}\rangle+\langle\bar{x}_{2}\rangle\nonumber \\
&=& -\langle\bar{x}_{3}\rangle\nonumber
\end{eqnarray}
where the last equality in Eqn \ref{Ci3wire} follows from the fact that summing over all edges E,
\begin{equation}\label{sumxbar}
\sum_{p=1}^{E}\langle\bar{x}_{p}\rangle=\sum_{p=1}^{E}\langle x_{p}\rangle-x_{00}=0
\end{equation}
by definition of $x_{00}$.

Eqn \ref{Fbc} then requires that
\begin{eqnarray}\label{Fcontinuity}
F_{1}(a_{1})&=&F_{2}(0)\\
F_{2}(a_{2})&=&F_{3}(0)\nonumber
\end{eqnarray}
It is straightforward to arrange these conditions once the Fs are computed from Eqns \ref{Fip}, \ref{Cp}, and \ref{Cout}.  Note that this leaves one overall undetermined constant, $F_{1}(0)$. We can always choose this constant such that the unions of the edge DL functions have zero ground state expectation value, $\langle 0|F|0\rangle =0$.  The Gs are computed in the same way using \ref{Gijp}, \ref{Dp}, and \ref{Dout}.  This allows us to use the simple expressions for the hyperpolarizabilities displayed in Eqns \ref{beta_ijk_simple} and \ref{gamma_ijkl_simple}.

\subsection{Star graphs}\label{sec:star}

Consider next a 3-star graph, as shown in the center panel of Figure \ref{fig:3graphs}.  If we write the three edges in the canonical form of Eqn \ref{Fip} and measure distance from the center vertex, we find from Eqn \ref{Cp} and Eqn \ref{Cout} that $C_{1}=-\langle\bar{x}_{1}\rangle$, $C_{2}=-\langle\bar{x}_{2}\rangle$, and $C_{3}=-\langle\bar{x}_{3}\rangle$ which is also equal to $\langle\bar{x}_{1}\rangle+\langle\bar{x}_{2}\rangle$ by Eqn \ref{sumxbar}. All of the $F_{x}^{(p)}$ are equal at the central vertex, and this is automatically satisfied if each of the F constants in Eqn \ref{Fi} are equal.  The value of this single constant may be set by demanding $\langle 0|F|0\rangle =0$, as noted above.

\begin{figure}\centering
\includegraphics[width=3.4in]{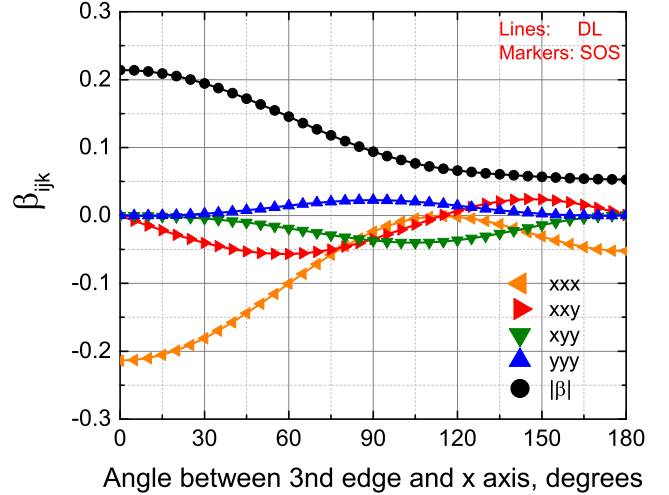}\\
\caption{Comparison of the Dalgarno-Lewis results for $\beta_{ijk}$ for the 3-star graph with those obtained using the full SOS for all tensor components. In the legend, $xxx\equiv\beta_{xxx}$, etc.}\label{fig:beta_tensors_vs_angles_DL_3star_degen}
\end{figure}

\begin{figure}\centering
\includegraphics[width=3.4in]{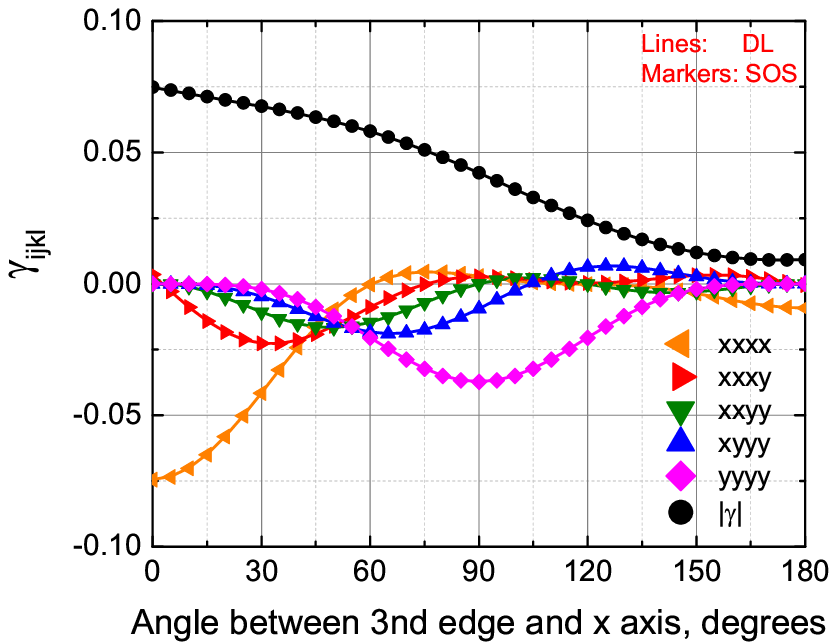}\\
\caption{Comparison of the Dalgarno-Lewis results for $\gamma_{ijkl}$ for the 3-star graph with those obtained using the full SOS for all tensor components. In the legend, $xxxx\equiv\gamma_{xxxx}$, etc.}\label{fig:gamma_tensors_vs_angles_DL_3star_degen}
\end{figure}

Figure \ref{fig:beta_tensors_vs_angles_DL_3star_degen} shows the results of using DL to calculate all of the tensor components of $\beta_{ijk}$ for a 3-star graph with two fixed edges of lengths $0.4$ and $0.2$, angles of $180$ and $90$ degrees relative to the x-axis, and one rotating edge of fixed length $a_{3}=0.6$, and compares them to those calculated by using the sum over states with thirty modes.  The DL method works with high precision.  Figure \ref{fig:gamma_tensors_vs_angles_DL_3star_degen} shows the calculations for $\gamma_{ijkl}$.  Note that the length scale is arbitrary, as these are intrinsic hyperpolarizabilities; the relative lengths determine the response of the structure.

The most remarkable thing about using DL for these sets of star graphs is that states $(4,5)$, $(10,11)$, etc. are degenerate in energy (where $n=0$ is the ground state), and the sum over states calculation requires identification of these states and diagonalization of the degenerate subspaces before it may be used.  The DL computations don't know about any of this--they just require the (nodeless) ground state wavefunction.  The DL operators, which implicitly sum up all of the contributions in perturbation theory from the excited states, automatically account for the degeneracies, as they followed from completeness of the unperturbed eigenstates.

\begin{figure}\centering
\includegraphics[width=3in]{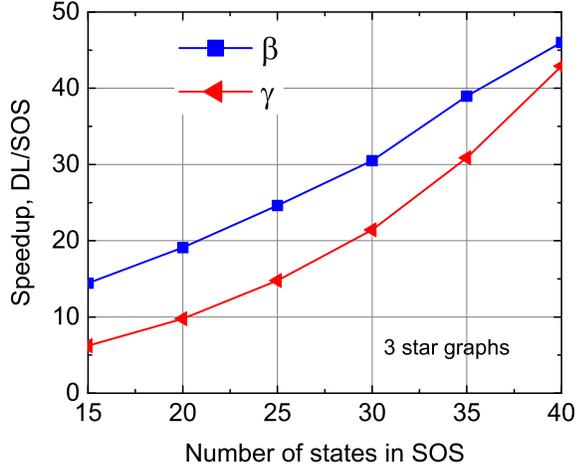}\\
\caption{Speedup of Dalgarno-Lewis computations over the sum over states.}\label{fig:speedup}
\end{figure}

The second remarkable feature of the DL summation technique is illustrated in Figure \ref{fig:speedup}, which shows the massive speedup of DL over the sum over states with large numbers of modes.
\begin{figure}\centering
\includegraphics[width=3.4in]{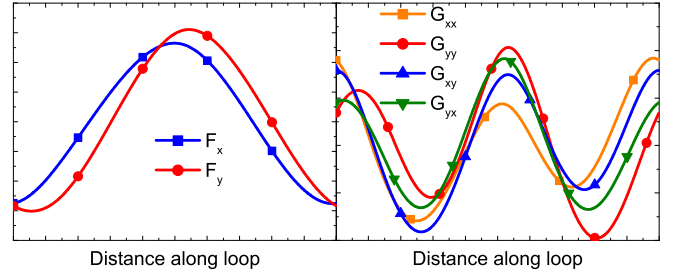}\\
\caption{Plots of the $F_{i}$ and $G_{ij}$ functions for the closed loop graphs, showing their periodicity, as expected.  Numerical values on the y axis have been suppressed for clarity.}\label{fig:FGloop}
\end{figure}

\subsection{Loops}\label{sec:loops}
Figure \ref{fig:3graphs}, right, shows a three-sided, closed loop graph.  The computation of $\beta_{ijk}$ and $\gamma_{ijkl}$ for this class of graphs has been extensively discussed\cite{shafe12.01}.  The ground state is nondegenerate, and is a constant, with zero energy.  The excited state spectrum is doubly-degenerate.  The eigenvalues are $k_{n}=2\pi n/L$, where $L$ is the perimeter of the loop.

The computation of the hyperpolarizabilities using DL may be accomplished by using Eqn \ref{Fip} but with a single edge and sequential calculation\cite{lytel12.01}.  For the x direction, say, we get
\begin{eqnarray}\label{Fsingle}
F(s)&=&F(0)+\frac{2m}{\hbar^2}\int_{0}^{s}\frac{ds'}{\phi(s')^{2}}\\
&\times&\left(\int_{0}^{s'}ds''\bar{x}(s'')\phi(s'')^{2}+ C\right)\nonumber
\end{eqnarray}
with a similar equation for the y component of F.  For a continuous wire graph, including a loop, we may use the sequential edge method of calculation\cite{lytel12.01}, where we measure distance s along the graph from an arbitrary starting point and change the value of $\bar{x}(s)$ in Eqn \ref{Fsingle} when we move from one edge to the next.  In this way, we can solve for the F and G functions for this graph by using Eqn \ref{Fsingle} and an equivalent one for G by selecting the correct boundary conditions.

For a loop graph, we demand that $F(L)=F(0)$, in order to make F periodic as the particle circulates the graph.  From Eqn \ref{Fsingle}, this requires that the first integration constant C be chosen so that the double integral on the right hand side vanishes at $s=L$.  We arrive at
\begin{eqnarray}\label{Fbc}
C&=&-\int_{0}^{L}\frac{ds'}{\phi(s')^{2}}\left(\int_{0}^{s'}ds''\bar{x}(s'')\phi(s'')^{2}\right)\nonumber \\
&\div&\int_{0}^{L}\frac{ds'}{\phi(s')^{2}}
\end{eqnarray}
for the first integration constant.  Finally, we note that the slopes are continuous, $F'(L)=F'(0)$, since the integral of $\bar{x}$ over the graph is zero by definition.

Similar remarks hold for the $G$ functions.  For example, for $G$ in the x-direction, we find
\begin{eqnarray}\label{Gsingle}
G(s)&=&G(0)+\frac{2m}{\hbar^2}\int_{0}^{s}\frac{ds'}{\phi(s')^{2}}\\
&\times&\left(\int_{0}^{s'}ds''\big[\bar{x}(s'')F(s'')-\langle 0|\bar{x}F|0\rangle\big]\phi(s'')^{2}+ D\right)\nonumber
\end{eqnarray}
where the first integration constant is given by
\begin{eqnarray}\label{Gbc}
D&=&-\int_{0}^{L}\frac{ds'}{\phi(s')^{2}}\\
&\times&\left(\int_{0}^{s'}ds''\big[\bar{x}(s'')F(s'')-\langle 0|xF|0\rangle)\big]\phi(s'')^{2}\right)\nonumber \\
&\div&\int_{0}^{L}\frac{ds'}{\phi(s')^{2}}\nonumber
\end{eqnarray}

Fig \ref{fig:FGloop} shows the two F functions and for G for the loop graph.  It is clear that the boundary conditions caused the functions to assume periodic behavior.

\subsection{Multiple star graph}\label{sec:7edge}
\begin{figure}\centering
\includegraphics[width=2.5in]{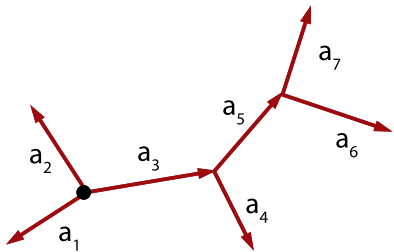}\\
\caption{Three prong graph with seven edges.}\label{fig:3prong}
\end{figure}

We can extend the general algorithm to more complicated structures, such as the three prong graph having seven edges (a star at each end, and a star connecting these), shown in the left panel of Figure \ref{fig:3prong}. For convenience, we can put the origin at the first star vertex, where edges $a_{1},a_{2},a_{3}$ are joined.  Using the algorithm in Eqn 4, we need to find seven first constants of integration, $C_{i},i=1-7$.  These are given by
\begin{eqnarray}\label{Ci3prong}
C_{1}&=&-\langle\bar{x}_{1}\rangle\nonumber \\
C_{2}&=&-\langle\bar{x}_{2}\rangle\nonumber \\
C_{3}&=&\langle\bar{x}_{1}\rangle+\langle\bar{x}_{2}\rangle\nonumber \\
C_{4}&=&-\langle\bar{x}_{4}\rangle\\
C_{5}&=&\langle\bar{x}_{1}\rangle+\langle\bar{x}_{2}\rangle+\langle\bar{x}_{3}\rangle+\langle\bar{x}_{4}\rangle\nonumber \\
C_{6}&=&-\langle\bar{x}_{6}\rangle\nonumber \\
C_{7}&=&-\langle\bar{x}_{7}\rangle\nonumber \\
\end{eqnarray}
Note that we could have used the algorithm to rewrite $C_{6}$ as
\begin{eqnarray}\label{constants_i}
C_{6}&=&\langle\bar{x}_{1}\rangle+\langle\bar{x}_{2}\rangle+\langle\bar{x}_{3}\rangle \\
&+&\langle\bar{x}_{4}\rangle+\langle\bar{x}_{5}\rangle+\langle\bar{x}_{7}\rangle\nonumber
\end{eqnarray}
by Eqn \ref{sumxbar}.

\begin{figure}\centering
\includegraphics[width=3.4in]{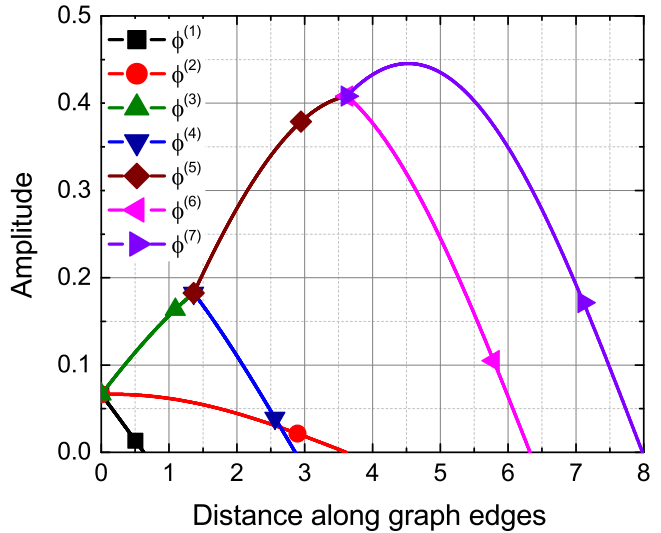}\\
\caption{Edge functions on three prong graph.  The x-axis denotes distance from the origin (black dot) in Fig 5.  The three vertices of this graph are shown as the intersection points of the edge functions.  Continuity of the edge functions and conservation of their derivatives at each vertex are evident in the Figure.}\label{fig:panel1_quad_3prong}
\end{figure}

Figure \ref{fig:panel1_quad_3prong} shows the edge functions along the graph and illustrates their continuity and the conservation of flux.  The edge lengths $a_1$ to $a_7$ are $0.63, 3.61, 1.36, 1.50, 2.26, 2.70$, and $4.36$, in arbitrary units, and their respective angles, in degrees, are $180, 60,0,60,0,36$, and $0$.  Edges 1,2,4,6,7 terminate at an infinite potential wall, and the edge functions vanish there, as may be seen in the Figure.

\begin{figure}\centering
\includegraphics[width=3.4in]{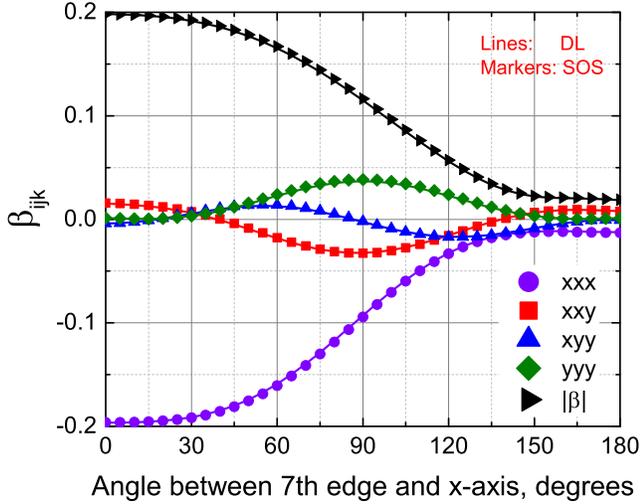}\\
\caption{First hyperpolarizability tensors, computed by DL (solid lines) and SOS (open lines) for the seven-edge graph described in the text.  Both methods agree to under a few hundreds of a percent.  The SOS calculation used 20 modes and executed ten times more slowly than the DL calculation.}\label{fig:panel2_quad_3prong}
\end{figure}

Figure \ref{fig:panel2_quad_3prong} shows that the SOS and DL results are identical using this formalism.  The DL calculations executed ten times faster than using 20 modes in the SOS calculation.  Moreover, the solution of the SOS calculation required careful computation of the amplitudes due to degeneracies in the graph.  As graphs get larger, degeneracies become very difficult to compute, and the SOS method is impractical.  The DL method does not require any information other than the ground state, which is nondegenerate.

\begin{figure}\centering
\includegraphics[width=3.4in]{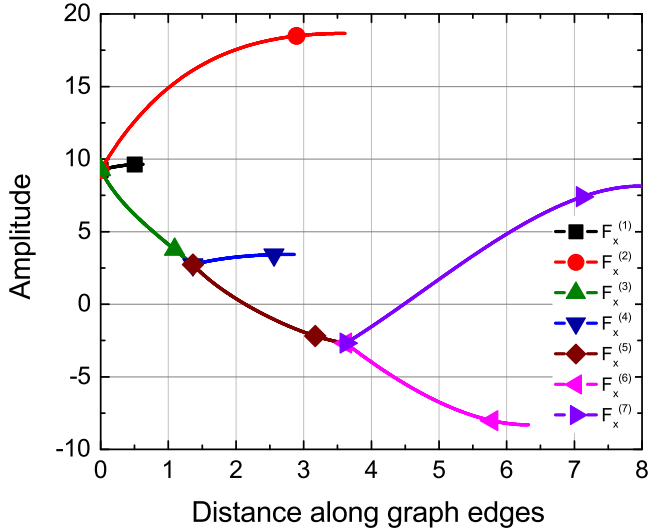}\\
\caption{$F_{x}^{(p)}$ functions for all graph edges using the same ordering scheme as used in Fig \ref{fig:panel1_quad_3prong} for the edge functions  Note that the DL edge functions are continuous at the vertices, and their derivatives are conserved at the vertices.}\label{fig:panel3_quad_3prong}
\end{figure}

\begin{figure}\centering
\includegraphics[width=3.4in]{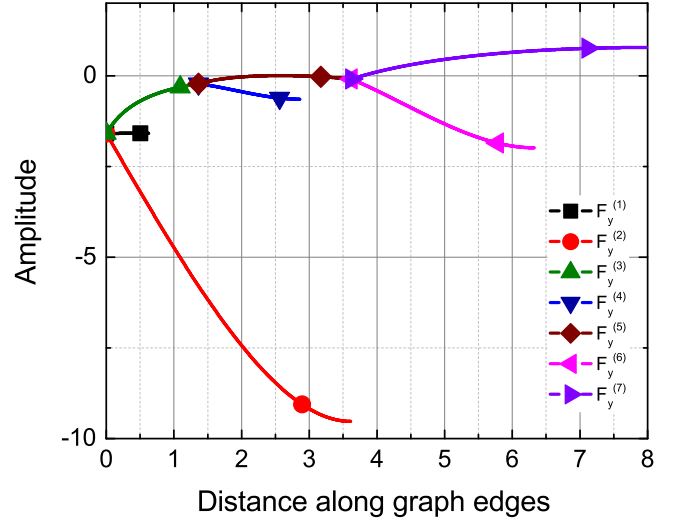}\\
\caption{$F_{y}^{(p)}$ functions for all graph edges using the same ordering scheme as used in Fig \ref{fig:panel1_quad_3prong} for the edge functions.  Note that the DL edge functions are continuous at the vertices, and their derivatives are conserved at the vertices.}\label{fig:panel4_quad_3prong}
\end{figure}

Figures \ref{fig:panel3_quad_3prong} and \ref{fig:panel4_quad_3prong} show the $F_{i}$ function along the x and y directions, respectively.  The figures show that the computation has produced edge DL functions that are continuous and have conserved slope at their vertices.

\section{Conclusions}\label{sec:end}
The computation of the optical nonlinearities of quasi-one dimensional systems, modeled by quantum graphs, has been adapted from the sum over states (SOS) to the Dalgarno-Lewis (DL) method and resulted in an execution time speedup of anywhere from a factor of $10-50$ for a simple star graph.  For large-scale Monte Carlo simulations normally taking hours with the SOS, the computation time is reduced to well under half an hour and often requires only a few minutes.  SOS requires solving the quantum graph for all of its eigenfunctions and energies, whereas DL requires only knowledge of the ground state and, for intrinsic nonlinearities, the first two energy levels.  A quantum graph with many edges has an exceedingly complex secular equation whose $M$ roots must be found for an SOS over $M$ states.  Often, the roots are so closely spaced that numerical solutions are inaccurate, and it is impossible, a priori, to even know which roots are degenerate, let alone what the degree of degeneracy is.  Subspaces of eigenfunctions with degenerate states must be diagonalized.  The time to sum over states for $\beta$ scales as $M^2$, while for $\gamma$, it scales as $M^3$.  The adaptation of the DL method for quantum graphs provides a fast, reliable way for researchers to solve quantum graph models that would otherwise be nearly intractable.

Another benefit of the DL method is that it requires only the ground state of the system.  This implies that one can, in principle, attempt to model structures with large response by testing various models of charge distribution in the ground state of a molecule.  In a real system, the Hamiltonian determines the exact eigenstates and spectra, and for a given set of boundary conditions, the ground state would be unique.  For modeling a system, one could mock up a ground state based upon intuition about how the charge is distributed in a test molecule, and then compute $\beta$ and $\gamma$ using the DL method.  The test molecule's actual Hamiltonian might not produce a ground state that exactly resembles the mocked up version, but the result should be meaningful.

The DL method has been solved for the dispersive case and is presented in a paper in this special issue.  The extension of the DL method for graphs to include dispersion is identical to the presentation here, except that the DL functions take a different form that depends on the optical frequencies and the linewidths.

\section{Acknowledgements}
SMM and MGK thank the National Science Foundation (ECCS-1128076) for generously supporting this work.


\end{document}